# On a dynamic ontic wave model of quantum collapse and measurement

Jason D. Runyan
Indiana Wesleyan University

## Statement of significance

This study addresses enduring problems in the foundations of quantum theory by introducing a novel, physically grounded model of quantum entities as unified, physically extended wavefields. The model derives the Born rule from localized physical interactions with an extended wavefield, reinterprets the Heisenberg uncertainty principle as a physical deformation constraint, and generates testable predictions relating energetic interaction to spatial localization. In doing so, it offers a unified realist and holist framework for quantum measurement and collapse that is consistent with existing experimental observations.

## Abstract

This work introduces a novel model of quantum entities as unified, physically extended wavefields, forming the basis for a testable realist, holist framework for quantum measurement and collapse. Unlike interpretations that postulate hidden variables, observer-induced effects, spontaneous stochastic collapse, or multiverse branching, this model derives the Born rule from the squared-amplitude structure of an extended wavefield undergoing localized, interaction-induced collapse. Central to the model is a reinterpretation of the Heisenberg uncertainty principle—not merely as a statistical or epistemic limitation, but as a dynamical relation between localized energetic interaction and wavefield localization. This framework yields testable predictions about how weak, intermediate, and strong quantum interactions modulate spatial localization—predictions consistent with existing experimental findings. The upshot is a unified, falsifiable alternative to standard interpretations, and a foundation for a broader research program in wavefield interaction mechanics.

For over a century, quantum mechanics has enjoyed unmatched success in modelling the behavior of entities such as electrons and photons. Yet despite this, its conceptual foundations remain unsettled and even controversial.

A universally accepted interpretation has proven elusive, largely because fully accounting for quantum phenomena seems to require interpretive assumptions that are speculative, counterintuitive, and/or philosophically contentious. To illustrate:

- The *Copenhagen interpretation* treats the physical location of quantum entities as indeterminate apart from measurement contexts, raising questions about the ontological status of unmeasured systems (cf. Bacciagaluppi and Valentini 2009; Faye 2019);
- *Many-worlds interpretations (Everettian)* posit a universal wavefunction with effectively branching quasi-classical worlds, raising questions about ontology, probability, and the status of unobservable branches (e.g., Wallace 2012);
- *Bohmian mechanics* introduces hidden variables and entails nonlocal dynamics, which challenge its compatibility with relativity (Bohm 1952; Bell 1987; Dürr et al. 1992; Goldstein 2021);
- *Objective collapse theories (e.g., GRW, rGRWf)* modify unitary evolution with stochastic collapse mechanisms, raising questions about the physical origin of the collapse process and, in some formulations, questions about energy conservation (e.g., Ghirardi et al. 1986; Tumulka 2006; Bassi et al. 2013); and
- *Epistemic and relational interpretations* (e.g., *QBism*) deny or qualify the view that the wavefunction represents an observer-independent physical entity or state, instead treating quantum states as agent-relative, relational, or informational (e.g., Rovelli 1996; Fuchs et al. 2014; Mermin 2014).

In this context, I propose a realist, holist alternative: the Dynamic Ontic Wave (DOW) model. This model retains the standard wavefunction formalism but interprets the wavefunction as representing the evolving structure of a physical wavefield, spatially extended in the sense that its amplitude, physical presence, and potential for interaction are distributed across space. It is holist because this wavefield is treated as a single unified quantum entity, with its own distinct physical tendencies, whose local manifestations depend on the structure and state of the whole. The physical structure of the wavefield evolves across space as modelled by the Schrödinger equation for massive quantum entities, or by other appropriate wave equations for different quantum systems. And it continues to do so unless locally and sufficiently disturbed by energetic interaction,[1] at which point the wavefield physically contracts—i.e., collapses—to a constricted location.

On this view, collapse is not spontaneous or observer-dependent. Nor is collapse triggered by a fundamentally spontaneous stochastic process. It is, rather, causally induced by sufficient localized energetic interaction—such as that involved in measurement or constraint. The occurrence of collapse is, thus, tied to determinate interaction conditions, even though the location to which the wavefield collapses remains probabilistic. Collapse is, then, an expression of how quantum entities (e.g., electrons) interact with other entities under energetic conditions. And the probability distribution of collapse locations is proportional to the squared amplitude of the wavefunction at the moment of collapse, thereby deriving the Born rule from the wavefield's spatial distribution under collapse conditions. In this way, though the time of collapse can be determinate, the location of collapse remains probabilistic.

[1] 'Energetic interaction' should be taken to include any physical process through which another system transfers energy or momentum to the wavefield, constrains its evolution, or imposes physically realized boundary conditions on its dynamics. Such interactions may weakly deform the wavefield, modulate its evolution, or, when sufficient, induce collapse (see Section 5).

The overarching implication is that quantum entities exhibit wave-like evolution and particle-like localization under differing conditions. The DOW model thus provides a unified ontology in which both wave- and particle-like states are modes of a single physical entity or system—eliminating the need for hidden variables, branching universes, spontaneous collapse events, or observer-dependent postulates. Instead, it offers a single physical ontology grounded in wavefield dynamics and localized interaction.

In what follows, I provide an initial development of the DOW model as a realist, holist theory which, more than being consistent with experimental observations, aims at a unifying explanation for seemingly paradoxical findings with the following virtues:

- *Ontic minimalism*—it postulates no hidden variables or parallel universes and provides a single ontology unifying both wave and particle behavior;
- *Dynamical unification*—it doesn't treat evolution and collapse as *ad hoc*, unrelated processes; a single framework explains evolution and collapse; and the Born rule is grounded in the spatial structure of an ontic wavefield;
- *Observer-independence*—collapse is triggered by physical interaction, not observation or knowledge; and
- *Causal grounding*—collapse results from localized energetic interaction under physical constraints, modelled via a reinterpretation of the Heisenberg uncertainty principle as a constraint rooted in wave mechanics (see Sections 2-4).

While these are conceptual virtues—not direct empirical confirmation—comparative theoretical analysis is the primary means to adjudicate between empirically equivalent interpretations—that is, theories that equally account for the relevant observations. That said, the DOW model also generates empirically testable predictions about energy-dependent localization which are consistent with existing experimental observations in weak, intermediate, and strong measurement regimes (e.g., Aharonov et al. 1988; Kocsis et al. 2011; Katz et al. 2006; Dressel et al. 2014; see Section 5). On this basis, it provides an organizing principle for a largely unexplored research program in wavefield interaction mechanics. I conclude by noting that whether the DOW model coheres with a full treatment of entanglement and can ultimately be given a relativistically covariant formulation remain open questions, and by outlining ways of approaching these questions (Section 6).

**1. Ontological commitments**

As introduced above, the DOW model retains the standard wavefunction formalism (see Section 2), but interprets the quantum wavefunction, $\psi(x,t)$, as representing the evolving physical structure of a spatially extended wavefield—the quantum entity in a wave-like mode of existence. So, the wavefunction represents the wavefield's changing ontic distribution rather than a merely epistemic probability assignment (cf. Harrigan and Spekkens 2010; Pusey et al. 2012). Quantum entities or systems are, in the absence of sufficient localized energetic interaction, wave-like, precisely in this ontological sense. There are no hidden particles, observer-induced effects, or multiple worlds in play.

Upon sufficient localized energetic interaction, quantum entities undergo collapse, which is a global, discontinuous contraction of a single unified entity to a constricted spatial location. This is not a smooth or unitary evolution, but a physically induced transition resulting from localized energetic interaction. This discontinuity marks a transition in the entity's physical state from spatially extended to sharply localized. In this sense, under such interaction conditions, quantum entities collapse into a particle-like state. This collapse isn't epistemic, nor is it triggered by a fundamentally stochastic process. It is a physical transformation of the wavefield brought about by sufficient localized energetic interaction.

On the DOW model, then, the wavefunction represents nothing other than the evolving physical structure and dynamical state of the quantum entity in its wave-like mode. Collapse is, therefore, a distinct, energy-dependent physical process, while the uninterrupted evolution of quantum entities is continuous and unitary—for example, as modelled by the Schrödinger equation for electrons and other massive quantum entities.

**2. Wavefunction evolution as a model for wave-like quantum states**

According to the DOW model, an electron is a physical, spatially extended wavefield whose evolution is represented by the wavefunction in the standard time-dependent Schrödinger equation (Schrödinger 1926):

$$i\hbar\frac{\partial\psi(x,t)}{\partial t} = \hat{H}\psi(x,t)$$

In this equation, $\psi(x,t)$ describes an evolving wavefield; that is, a wavefield whose physical structure and dynamical state are changing over time (cf. Wallace and Timpson 2010; Albert and Ney 2013; Ney 2021), with its amplitude corresponding to the wavefield's local relative concentration, and its phase to the wavefield's dynamical structure, including momentum and current structure where such quantities are well-defined. $\hat{H}$ is the Hamiltonian operator associated with the total energy (kinetic and potential) of the quantum entity.

The DOW model, thus, treats the wavefunction not as a mere mathematical abstraction or computational tool, but as a mathematical representation of the evolving physical structure and dynamical state of a quantum wavefield. This equation models the continuous, unitary, and determinate propagation of a quantum wavefield across space and time outside of collapse. No additional hidden parameters are required to account for this evolution or for the quantum state represented by the wavefunction.

### 3. Interaction-induced wave contraction: a discontinuous transition to a particle-like state

A central innovation of the DOW model is its account of how, when, and where wave contraction—or collapse—occurs. In this model, collapse is a physical process triggered by sufficient localized energetic interaction. As I will formally, though generally, specify in Section 5, this is an interaction in which localized energetic interaction reaches or exceeds a collapse threshold. When such an interaction occurs—for example, as a result of localized energy transfer during contact with a detector or localized constraint by a physical boundary—the wavefield undergoes a discontinuous, non-unitary contraction from an extended state to a spatially localized one.

Accordingly, the timing of collapse is determinate: it occurs under conditions of sufficient localized energetic interaction.[2] The location of the collapse is, however, probabilistic, with the probability distribution determined by the physical structure of the wavefield at the time of this interaction as represented by the wavefunction $\psi(x,t)$, and as modelled, for example, by the Schrödinger equation (see Section 2).

Collapse is formally modelled as the application of a collapse operator $\hat{C}_{x_0}$, where $x_0$ is the localization point realized in accordance with the probability distribution determined by the wavefunction:

$$\psi_{\text{collapsed}}(x) = \frac{\hat{C}_{x_0}\,\psi(x)}{\|\hat{C}_{x_0}\,\psi(x)\|}$$

[2] While—in keeping with standard formalizations—I will treat collapse as instantaneous here, the DOW model doesn't depend on whether collapse is truly instantaneous or occurs over a very short interval (see Section 6). This distinction is immaterial to the structure and explanatory aims of the model, which are compatible with the possibilities currently left open by empirical evidence concerning the temporal dynamics of collapse.

The operator may be represented—in a limiting idealization—as a Dirac delta function or—more physically—as a sharply peaked Gaussian centered at $x_0$ (Ghirardi et al. 1986). The denominator normalizes the state after collapse by dividing by the norm of the unnormalized collapsed state, ensuring that the total probability remains 1—thereby reflecting the global reconfiguration of a unified quantum entity.[3]

In the DOW model, collapse occurs at a definite time $t_0$—the time of collapse-inducing energetic interaction. But it is localized at position $x_0$ probabilistically, with the probability density for localization at $x_0$ given by the squared amplitude of the wavefunction—as determined by the appropriate dynamics for the entity in question (e.g., as modelled by the Schrödinger equation)—at the time of collapse-inducing interaction:

$$P(x_0) = |\psi(x_0, t_0)|^2$$

This formalism marks a sharp departure from interpretations in which collapse is treated as spontaneous, fundamentally stochastic in occurrence (e.g., GRW), observer- or agent-relative (e.g., QBism), or otherwise left physically and ontically undefined. While other models have proposed physical thresholds for collapse—most notably based on gravitational self-energy (Diósi 1989; Penrose 1996)—the DOW model identifies a threshold grounded in localized energetic interactions and the kinematic constraints of wave mechanics (see Section 5). And this is a pivotal explanatory strength of the model. For, by treating collapse as a well-defined, mathematically modelled physical process, not only is the Born rule naturally grounded in the geometry of the evolving wavefield (see Section 4), but a new basis is provided for the strategic manipulation of this evolution (see Section 5).

## 4. Collapse probabilities and the Born rule

Given the proposed model, the Born rule isn't introduced axiomatically, postulated, or otherwise imposed. Instead, given:

(1) the wavefunction represents an evolving spatially extended wavefield, and
(2) collapse is a localized contraction of this wavefield occurring at a determinate interaction time,

---

[3] The DOW model does not require a specific form for the collapse operator beyond the assumption that it is localized in space. This is broadly analogous to the treatment in leading collapse models: GRW, for example, assumes a Gaussian collapse operator with a fixed localization width, while CSL introduces a continuous stochastic process but similarly employs a spatially localized form. The DOW model allows the operator to be idealized as a delta function or more realistically modelled as a sharply peaked Gaussian. Since the key explanatory aim here is to account for the Born rule via interaction-induced collapse, the precise dynamical form of the operator is not essential to the current argument and is left open.

the distribution of collapse locations—which the Born rule describes—can be derived from the wavefield's spatial intensity distribution at that time.[4] In this way, within the DOW model, the Born rule is derived from the physical structure of a wavefield and its interaction-induced contraction: the probability density for localization at $x_0$ is determined by the squared amplitude of the wavefield at $x_0$ at the time of collapse $t_0$.

To illustrate:

Given quantum entities with mass evolve as modelled by the Schrödinger equation (see Section 2), and collapse at a determinate interaction time $t_0$ via a localized operator $\hat{C}_{x_0}$ (see Section 3), the probability density for collapse centered at $x_0$, $P(x_0, t_0)$, can be modelled as:

$$P(x_0) \propto \langle \psi(t_0) | \hat{C}^{\dagger}_{x_0} \hat{C}_{x_0} | \psi(t_0) \rangle$$

For a sharply peaked operator $\hat{C}_{x_0}$, reflecting the localized nature of collapse at a definite time, the probability density is proportional to the local squared amplitude:

$$P(x_0) \propto |\psi(x_0, t_0)|^2$$

To obtain a proper probability density while preserving the wavefield as a single normalized ontic whole, this expression must be normalized over space, which can be done by dividing it by the total probability across all space:

$$P(x_0) = \frac{|\psi(x_0, t_0)|^2}{\int |\psi(x, t_0)|^2 \, dx}$$

Since the wavefunction evolves unitarily prior to interruption, the integral is normalized to unity:

$$\int |\psi(x, t_0)|^2 \, dx = 1$$

[4] The use of the squared amplitude, rather than the amplitude itself, reflects the wavefield's local physical intensity in a wave-like state. Thus, just as the intensity of a classical wave is proportional to the square of its amplitude, the probability of an interaction-induced collapse of a wavefield in its wave-like modality arises from the wavefield's local intensity at the moment of this interaction, which is likewise proportional to the squared amplitude.

Thus, the collapse location probability simplifies to:

$$P(x_0) = |\psi(x_0, t_0)|^2$$

And this is precisely the Born rule: the probability density for finding the quantum entity at location $x_0$ at the time of collapse $t_0$ is given by the squared amplitude of the wavefunction at that location and time (Born 1926).

In this way, given the DOW model, the Born rule follows formally from a minimal number of physical commitments—each continuous with standard quantum formalism and experimental observation—without requiring observers, *ad hoc* measurement postulates, or branching worlds. Experimentally observed collapse location distributions, as described by the Born rule, are thus deducible from:

- Collapse occurring at the time of a collapse-inducing physical interaction (i.e., a determinate time);
- Collapse location being probabilistic; and
- The probability density for collapse location being determined by the wavefield's local intensity (which, in wave mechanics, is proportional to the squared amplitude).

Accordingly, repeated measurements of collapse location will yield statistical patterns that reflect the wavefield's evolving physical structure and dynamical state—as represented by the wavefunction's amplitude structure—at the moment of interaction-induced collapse.

**5. Wavefield interaction mechanics**

If wavefunctions represent evolving spatially extended wavefields, and these wavefields collapse as a result of sufficient localized energetic interaction (cf. Box 1), then this raises the possibility of an energetic threshold for interaction-induced collapse. In this case, collapse occurs only when an interaction—including a measurement or other localized physical constraint—results in sufficient energy transfer to transition a wavefield into a localized state. In other words, collapse occurs when localized energetic interaction reaches the threshold required to constrict the wavefield into a localized state. While interactions below this threshold will not induce collapse, they may still produce measurable physical modulations in wavefield evolution.

The implication is, thus, that sub-threshold disturbances should be treated as physical effects—not merely heuristically, abstractly, or epistemically. And this, in turn, yields a rigorous, theoretically grounded approach to systematically modelling and strategically investigating weak and intermediate measurement effects as energy-dependent wavefield disturbances.

A natural hypothesis, then, is that the degree of disturbance—up to the point of collapse—increases as a function of localized energetic interaction, and that these effects unfold in opposition to the wavefield's intrinsic tendency toward continuous, unitary evolution. In other words, the wavefield tends to maintain coherent evolution rather than localize unless there is sufficient energetic interaction to trigger a transition to a localized state. Therefore, modelling this physical process requires capturing both the energy-dependent disturbance of the wavefield and the wavefield's tendency toward unitary evolution up to the collapse threshold.

This hypothesis provides a physical explanation for an uncontroversial feature of quantum measurement: greater positional precision requires greater momentum spread and a corresponding increase in the energetic scale of the interaction. It also yields a novel ontological reinterpretation of the Heisenberg uncertainty principle (Heisenberg 1927; Kennard 1927; Robertson 1929):

$$\Delta x \, \Delta p \geq \frac{\hbar}{2}$$

Given the DOW model, position uncertainty $\Delta x$ corresponds to the physical spread of a wavefield, while momentum uncertainty $\Delta p$ reflects the momentum spread that results from wavefield constriction. This momentum spread, thus, corresponds to the energy required to constrict a wavefield to a given spatial width. Up to the point of collapse, the greater the spatial constriction, the greater the resulting momentum spread and the greater the energy required. This relation between $\Delta p$ and the energetic scale of localization—like the Heisenberg uncertainty principle itself—is rooted in the Fourier structure of wave mechanics: the narrower a wave is in position, the broader its momentum distribution. This gives the principle an ontological interpretation that, while novel, returns it to its roots in wave mechanics.

The uncertainty principle, then, is not merely a statistical or epistemic constraint. Rather, it reflects the intrinsic structure of a wavefield and the energetic cost of constricting it, thereby disrupting its continuous evolution. Specifically, the relation between position and momentum demonstrates that momentum spread, and thus the energy required for localization, must increase inversely with spatial localization width. And solving for $\Delta p$, we get:

$$\Delta p \geq \frac{\hbar}{2\Delta x}$$

So, then, as the quantum entity becomes more localized, its momentum uncertainty increases, reflecting the greater momentum spread that results from—and the corresponding energy scale required for—constricting the wavefield to a narrower spatial region.

While localization may occur through various means of energetic interaction—by way of example—for massive quantum entities modeled nonrelativistically, the relevant localization energy scale can be estimated using the kinetic energy relation:

$$E_k = \frac{p^2}{2m}$$

Substituting the uncertainty-bound value for $\Delta p$ into this expression gives an idealized minimum kinetic energy scale required to constrict a wavefield to within spatial width $\Delta x$:

$$E_{\text{localization}}(\Delta x) \gtrsim \frac{(\Delta p)^2}{2m} \geq \frac{\hbar^2}{8m(\Delta x)^2}$$

Thus:

$$E_{\text{localization}}(\Delta x) \gtrsim \frac{\hbar^2}{8m(\Delta x)^2}$$

Under these idealized nonrelativistic conditions, this defines the minimum kinetic energy required to constrict a wavefield's spatial extent to a width $\Delta x$. Consequently, given the DOW model, this energy-localization relation identifies a natural lower-bound estimate of the collapse threshold scale:

$$E_{\text{collapse}}(\Delta x) \gtrsim \frac{\hbar^2}{8m(\Delta x)^2}$$

So, if the localized energy transferred in an interaction equals or exceeds the relevant collapse threshold, localization and thus wavefield collapse occurs. If, however, the transferred energy is below this, the wavefield may be physically disturbed (if it is sufficient to alter its spatial distribution) without full collapse. That is, its amplitude and/or phase may be altered, but it will continue to evolve without collapse under the appropriate dynamics. In this way, sub-threshold energy interactions define a non-collapse modulation range.

To model this, let $\Delta x_0$ be the initial wavefield width and $\Delta x_c$ the finite localization width reached in collapse. We can then introduce a schematic disturbance or deformation function for energy transfer below the collapse threshold, $0 < E < E_{\text{collapse}}(\Delta x_c)$:

$$\Delta x(E) = \Delta x_c + (\Delta x_0 - \Delta x_c)\left(1 - \left(\frac{E}{E_{\text{collapse}}(\Delta x_c)}\right)^{\gamma}\right)$$

where $\gamma > 0$ controls the deformation curvature ($\gamma = 1$ gives linear narrowing; $\gamma > 1$ gives faster initial narrowing; $0 < \gamma < 1$ gives slower initial narrowing);[5] $\Delta x(E) \to \Delta x_0$ as $E \to 0$; and $\Delta x(E) \to \Delta x_c$ as $E \to E_{\text{collapse}}(\Delta x_c)$, marking collapse to a finite localization width.

This provides a general schematic approach for modelling how a wavefield's effective spatial width decreases under sub-threshold interactions and approaches localization as the collapse threshold is approached.

To reinforce the physical plausibility of this model, we may minimally formalize the relationship between localized energy transfer and global wavefield deformation.

Let $\psi_0(x,t)$ represent the undisturbed wavefield evolving under the appropriate dynamics, and let $\psi_E(x,t)$ represent the wavefield after localized energy transfer $E$. Assuming both states are normalized, the global deformation induced by this interaction can then be represented by the loss of overlap between the undisturbed and disturbed wavefield states:

$$\mathcal{D}(E,t) = 1 - |\langle \psi_0(t) \mid \psi_E(t) \rangle|^2$$

Collapse occurs when this global deformation exceeds a critical threshold:

$$\mathcal{D}(E,t) \geq D_c, \qquad 0 < D_c < 1$$

Below this threshold, localized interaction may alter the wavefield's amplitude and/or phase while preserving evolution outside of collapse. At or above this threshold, the deformation is sufficient to induce global contraction of the wavefield *qua* whole to a localized state.

This formalism expresses a core premise of the DOW model: localized energy transfer can deform the wavefield globally as the wavefield is a single, unified spatially extended entity. Collapse occurs not through stochastic or observer-dependent processes, but when localized energy transfer exceeds the relevant threshold, thereby deforming the wavefield

[5] While $\gamma$ is a free parameter in this initial formulation, it may be empirically constrained by future experiments examining wavefield localization under finely graded interaction energies.

as a whole. And—as noted at the outset—this deformation is constrained by the uncertainty relation:

$$\Delta x(E) \cdot \Delta p(E) \geq \frac{\hbar}{2}$$

where $\Delta p(E)$ quantifies the increased momentum uncertainty arising from constriction of the wavefield to width $\Delta x(E)$, and thus the corresponding energy scale required for doing so. And this is simply the Heisenberg uncertainty principle reinterpreted as a physical coherence constraint rather than merely a statistical or epistemic limitation. In effect, the principle thus models a physical boundary on modulation, rather than a mere statistical bound. It relates the degree of localization the wavefield sustains up to collapse to the corresponding energetic scale of interaction.

In sum, given the DOW model, the uncertainty principle formalizes the relation between a wavefield's spatial width and the kinetic energy required to constrict it. Collapse is, then, no longer an unexplained discontinuity, but the natural limit of continuous, energy-constrained deformation of a unified wavefield. The model thereby unifies wave-like and particle-like quantum behavior as modes of a single, energy-dependent physical process.

Moreover, if weak, intermediate, and strong measurements correspond to energetic interactions of increasing magnitude—resulting in increasingly localized wavefields—then the uncertainty principle provides a predictive framework for relating these energetic interactions to localization. This is because it enables the formal derivation of energy-localization relationships.

To illustrate, using an electron's rest mass ($m_e \approx 9.11 \times 10^{-31}$ kg) along with the equation:

$$E = \frac{\hbar^2}{8 m_e \Delta x^2}$$

We obtain the predicted energy-localization relationships under idealized conditions shown in Table 1.

**Table 1.** *Predicted minimum kinetic energy scale required to localize an electron wavefield to a given spatial scale in idealized nonrelativistic conditions, based on the uncertainty relation.*[6]

| **Localization Scale** $\Delta x$ (nm) | **Predicted Energy** (eV) |
|---|---|
| 1 | ≈0.009524 |
| 0.1 | ≈0.9524 |
| 0.01 | ≈95.24 |
| 0.001 | ≈9524 |

These predictions are consistent with the broad energy-localization scaling observed in electron measurement techniques. Scanning tunneling microscopy and photoemission studies involve electron scale energies on the order of eV and can probe localization at sub-nanometer scales (e.g., Binnig and Rohrer 1987). Transmission electron microscopy uses electron beams in the ~100–300 keV range and resolves spatial features at the picometer to sub-angstrom scales (e.g., Egerton 2009). While resolution depends on experimental geometry, interaction type, coherence, detector response, and reconstruction methods, what observations have been made are consistent with the DOW model's general premise: finer localization requires stronger localized energetic interaction all the way up to collapse, and, short of full collapse, weaker interactions produce less localization.

While this illustration is developed for electrons—where wavefield deformation can be modelled nonrelativistically using Schrödinger dynamics—this is for ease of initial conceptualization. The underlying principle of energy-dependent modulation may generalize across quantum entities. This includes massless particles like photons, though extending the model in this way would require a relativistic or field-theoretic formulation (see Section 6). Notably, because massless entities follow a linear energy-momentum relation ($E=pc$), a relativized extension would be expected to yield energy-localization scaling closer to $1/\Delta x$, rather than the $1/\Delta x^2$ scaling derived for massive entities modelled nonrelativistically. Nevertheless, consistent with this broader prediction, weak measurement experiments involving photons—such as those by Kocsis et al. (2011), Lundeen et al. (2011), and Piacentini et al. (2016)—indicate that weak energetic interactions can alter wave behavior without inducing collapse. This observed relationship between interaction strength and localization behavior is therefore consistent with the DOW model's broader premise that energetic interaction, when sufficient to disturb a wavefield, affects quantum localization. This suggests that the underlying energy-localization relation may be generalizable, thereby motivating a relativized formulation (see Section 6).

[6] These predictions reflect idealized, lower-bound energy thresholds based on simplified wavefield assumptions and isolated system dynamics. In practical experimental settings, collapse may require somewhat higher energies due to environmental coupling, asymmetries in system configuration, detector geometry, or dissipative effects. Consequently, observed localizations will often—expectedly—involve energies well above the theoretical minimum without contradicting the model.

Thus, the DOW model not only reinterprets the uncertainty principle in terms of wavefield dynamics but also predicts an energy-localization relationship consistent with a range of empirical observations. This strengthens the model's plausibility and positions it as an empirically motivated, predictive framework for quantum measurement and collapse. It, likewise, opens a principled approach to investigating how weak, intermediate, and strong interactions modulate the behavior of quantum entities, whether through sigmoid-like, exponential, linear, or power-law relationships, and whether such modulations involve thresholds. More broadly, this formalism initiates a new research program: wavefield interaction mechanics.

Through investigations of how localization may be modulated through energy tuning, this program may help organize research into controlled interactions with quantum entities—a pathway toward refined measurement protocols, real-time feedback on wavefield modulation, and improved strategies for managing measurement back-action in quantum information systems. In this way, the DOW model may provide an organizing principle for interpreting and advancing experimental work indicating that energetic interactions shape wavefield amplitude, phase, and temporal evolution (Itano et al. 1990; Kwiat et al. 1995; Katz et al. 2006; Korotkov and Jordan 2006; Devoret and Schoelkopf 2013; Dressel et al. 2014; Muhonen et al. 2014; Campagne-Ibarcq et al. 2016), and thereby shape the probability landscape of quantum collapse.

## 6. Conclusion: Ontic waves, entanglement, and relativization

There is historical precedent for the kind of conceptual shift proposed in this work.

Einstein advanced physics by proposing that Planck's quantization of energy—rather than being merely a mathematical strategy—reflects a physical property of light, suggesting that what had been treated primarily as a wave also exists as discrete quanta (Planck 1901; Einstein 1905).

In a similar spirit, the DOW model proposes that the wavefunction—often treated merely as a computational tool—reflects the evolving physical structure and dynamical state of quantum entities, suggesting that these entities exist as spatially extended wavefields. And just as recognizing the particle-like modality of light led to major progress, it may now be fruitful to fully acknowledge the wave-like modality of quantum entities—accepting that they behave as waves in their evolution and as particles upon sufficient energetic interaction.

In this case, wave-particle duality is not a contradiction or heuristic, nor need it remain a puzzle or paradox, but reflects two energy-dependent modalities of an ontically unified whole. And conceiving of this duality as ontologically unified in this way presents a shift in thinking about quantum behavior that may inform the strategic design of weak and moderate interactions to intentionally shape the probability distribution of collapse locations at specific future times.

In short, by treating the wavefunction as a model for an evolving physical wavefield, and collapse as a physically constrained, energy-dependent, and quantifiable process, the DOW model lays the foundation for a broad research program—wavefield interaction mechanics—that may both explain and guide controlled manipulation of quantum systems.

The DOW model is, therefore, not only realist, but holist. It treats quantum entities and systems as spatially extended wavefields whose local manifestations depend on the structure and state of the whole. This is consonant with priority monism (Schaffer 2010) and quantum holism (Howard 1985; Teller 1986; Healey 1991; Maudlin 2007; Ismael and Schaffer 2020), according to which wholes may be ontologically and/or explanatorily prior to their parts or aspects; and, in keeping with this, quantum systems may express distinct tendencies, or dispositions, that are irreducible to interactions between independently specifiable parts. In this way, the DOW model can be seen as proposing a physical and practical application of such approaches. Thus, according to this model, localized interaction affects the wavefield as a whole because the wavefield is an ontically unified extended entity, not an aggregate, and treating it as such can inform controlled interactions with quantum entities.

This holist realist approach to quantum phenomena also suggests a principled way of addressing entanglement and relativization. For, given this general approach, entanglement correlations needn't be treated as signals transmitted between an aggregate of local particles. Rather, because entangled systems cannot be fully specified by assigning independent states to their spatially separated parts, they are natural candidates for the kind of holist treatment of quantum entities proposed here. Accordingly, it is worth exploring whether entanglement reflects the behavior of a whole, non-aggregative wavefield system under localized interaction. Further, a relativized formulation of the DOW model might help distinguish global state reconfiguration from such things as superluminal signaling. By treating this reconfiguration as holistic rather than dynamically communicative, such a model could preserve empirically supported no-signaling (cf. Bell 1964; Tumulka 2006) while treating collapse as a holistic transition in the state of a unified quantum system, rather than as a local transmission between separate aspects. Developing this idea covariantly would require specifying how localized interactions constrain non-aggregative wavefield states without treating the resulting global reconfiguration as a signal transmitted through space (e.g., Dorato 1996; Tumulka 2006). Whether this can be done remains an open question, but the holist realism of the DOW model supports a promising path forward—or at least one worth exploring.

Nevertheless, many otherwise puzzling aspects of quantum behavior appear to fall into place once we shift to thinking of quantum entities as spatially extended, ontically unified wavefields. Entanglement and relativization may require further extensions of this framework, but the underlying holist realist approach to wavefields points a way forward. Perhaps this possibility has remained obscured because quantum phenomena have too often been approached through reductionistic assumptions rather than by considering holistic possibilities.

**Box 1.** The double-slit experiment & interaction-induced collapse.

In a canonical double-slit experiment, quantum entities such as electrons or photons pass through a barrier (B) containing two slits. Their positions are subsequently detected on a screen (S) placed behind the barrier (e.g., Grangier et al. 1986; Tonomura et al. 1989).

**1. When no measurement occurs at (B)**, and detection takes place only at (S), an interference distribution is produced. In the DOW model, this is because the wavefield, though modulated by the two-slit barrier, remains coherently extended across both paths. Its amplitude and phase structure evolve through the two slits, producing an interference pattern reflective of its physical structure and dynamical state.

**2. When a measuring device at or near (B)** interacts with the wavefield at or near the slits, the interference pattern disappears if the interaction transfers sufficient localized energy to induce collapse. In that case, coherent extension is disrupted, and the wavefield localizes near one slit before reaching the screen (cf. Scully et al. 1991; Englert 1996; Zurek 2003). Thus, the subsequent distribution at (S) no longer reflects the interference produced by coherent propagation through two slits.

On this interpretation, the behavior of the quantum entity does not require observer-dependent collapse, hidden variables, or branching worlds. It simply reflects whether the wavefield remains coherently extended through both slits or is locally collapsed by sufficient energetic interaction before reaching the screen. In either case, the probability distribution of localization is grounded in the squared amplitude of the wavefield at the moment of collapse. It is just that the collapse-inducing measurement at or near the two-slit barrier constrains transmitted localization outcomes to one slit or the other, while the wavefield's extended structure at that moment determines the probability distribution over these outcomes. Moreover, whether the wavefield is transmitted, absorbed, or scattered at the barrier depends on the interaction between the wavefield's extended structure, the structure of the barrier, and the location and strength of any collapse-inducing interaction.

The same general principle may extend to explain delayed-choice, quantum eraser, and weak-measurement variants (cf. Scully et al. 1991; Aharonov et al. 1988; Kocsis et al. 2011): wavefields evolve coherently unless locally disturbed by energy transfer at or above the collapse threshold, while sub-threshold interactions may modulate amplitude and/or phase without inducing full collapse. Thus, what ultimately explains this behavior is the intrinsic tendency of wavefields toward coherent evolution and their interactions with other entities. Photon cases would require the appropriate electromagnetic or field-theoretic formulation, whereas the nonrelativistic formalization in Section 5 applies most directly to massive particles such as electrons.

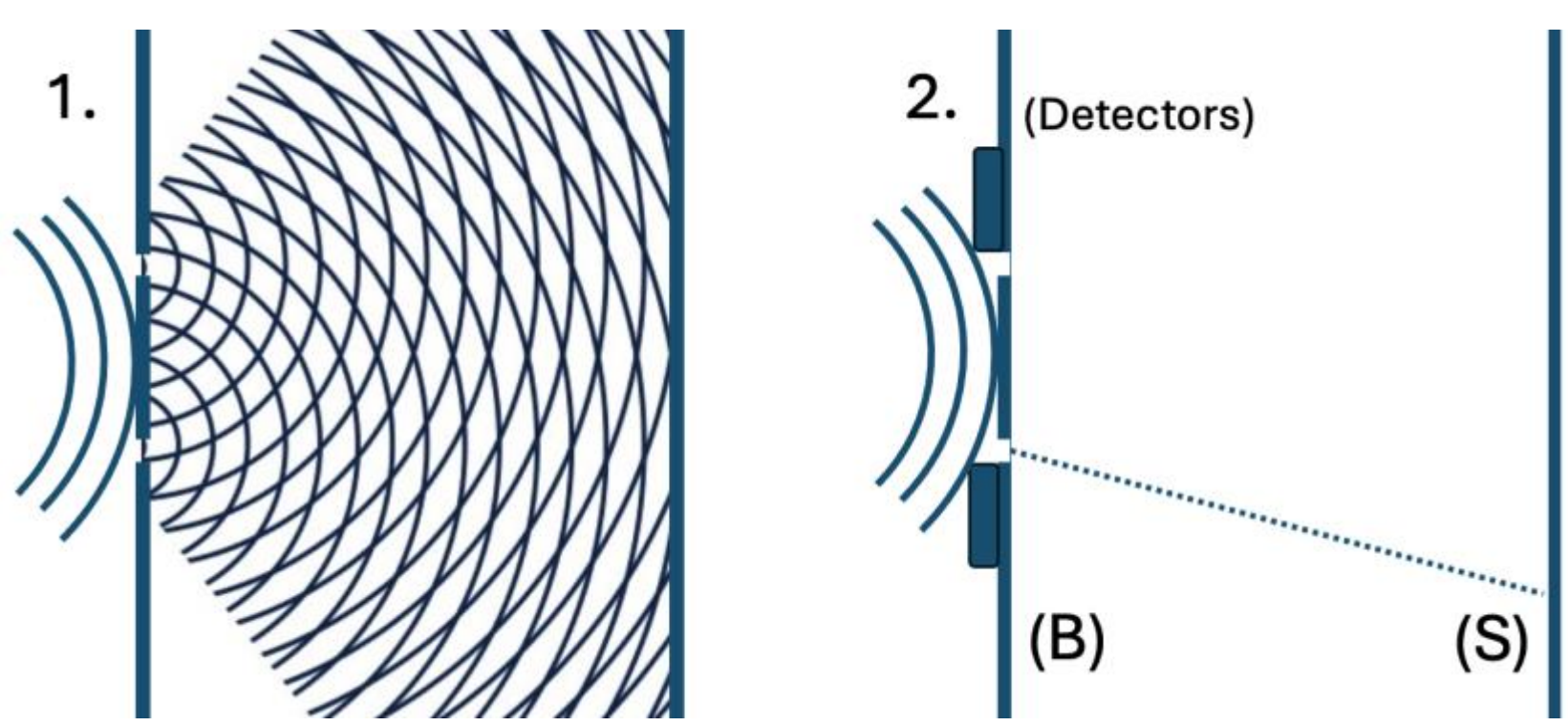

**References**

Aharonov, Y., Albert, D. Z. & Vaidman, L. How the result of a measurement of a component of the spin of a spin-1/2 particle can turn out to be 100. *Phys. Rev. Lett.* **60**, 1351–1354 (1988).

Bacciagaluppi, G. & Valentini, A. *Quantum Theory at the Crossroads: Reconsidering the 1927 Solvay Conference*. Cambridge Univ. Press (2009).

Bassi, A., Lochan, K., Satin, S., Singh, T. P., & Ulbricht, H. Models of wave-function collapse, underlying theories, and experimental tests. *Rev. Mod. Phys.* **85**, 471–527 (2013).

Bell, J.S. *On the Einstein–Podolsky–Rosen paradox*. Physics **1**, 195–200 (1964).

Bell, J.S. *Speakable and Unspeakable in Quantum Mechanics*. Cambridge University Press (1987).

Binnig, G. & Rohrer, H. Scanning tunneling microscopy—from birth to adolescence. *Rev. Mod. Phys.* **59**, 615–625 (1987).

Bohm, D. *A Suggested Interpretation of the Quantum Theory in Terms of "Hidden" Variables I and II*. Phys. Rev. **85**, 166–193 (1952).

Born, M. *Zur Quantenmechanik der Stoßvorgänge*. Zeitschrift für Physik **37**, 863–867 (1926).

Campagne-Ibarcq, P. et al. *Observing quantum state diffusion by heterodyne detection of fluorescence*. Phys. Rev. X **6**, 011002 (2016).

Devoret, M.H. & Schoelkopf, R.J. *Superconducting circuits for quantum information: An outlook*. Science **339**, 1169–1174 (2021).

Diósi, L. Models for universal reduction of macroscopic quantum fluctuations. *Phys. Rev. A* **40**, 1165–1174 (1989).

Dorato, M. On becoming, relativity, and nonseparability. *Philos. Sci.* **63**, 585–604 (1996).

Dressel, J. et al. *Colloquium: Understanding quantum weak values*. Rev. Mod. Phys. **86**, 307–316 (2014).

Dürr, D., Goldstein, S. & Zanghì, N. Quantum equilibrium and the origin of absolute uncertainty. *J. Stat. Phys.* **67**, 843–907 (1992).

Egerton, R. F. Electron energy-loss spectroscopy in the TEM. *Rep. Prog. Phys.* **72**, 016502 (2009).

Einstein, A. *On a Heuristic Point of View about the Creation and Conversion of Light*. Ann. Phys. **17**, 132–148 (1905).

Englert, B.-G. Fringe visibility and which-way information: An inequality. *Phys. Rev. Lett.* **77**, 2154–2157 (1996).

Faye, J. *Copenhagen Interpretation of Quantum Mechanics*. In: Zalta, E.N. (ed.) The Stanford Encyclopedia of Philosophy (2019).

Fuchs, C. A., Mermin, N. D. & Schack, R. An introduction to QBism with an application to the locality of quantum mechanics. *Am. J. Phys.* **82**, 749–754 (2014).

Ghirardi, G.C., Rimini, A., Weber, T. *Unified dynamics for microscopic and macroscopic systems*. Phys. Rev. D **34**, 470 (1986).

Goldstein, S. Bohmian mechanics. In Zalta, E. N. (ed.) *The Stanford Encyclopedia of Philosophy* (Fall 2021 edn.). Stanford Univ. (2021).

Grangier, P., Roger, G. & Aspect, A. Experimental evidence for a photon anticorrelation effect on a beam splitter: A new light on single-photon interferences. *Europhys. Lett.* **1**, 173–179 (1986).

Harrigan, N. & Spekkens, R. W. Einstein, incompleteness, and the epistemic view of quantum states. *Found. Phys.* **40**, 125–157 (2010).

Healey, R. A. Holism and nonseparability. *J. Philos.* **88**, 393–421 (1991).

Heisenberg, W. Über den anschaulichen Inhalt der quantentheoretischen Kinematik und Mechanik. *Z. Phys.* **43**, 172–198 (1927).

Howard, D. Einstein on locality and separability. *Stud. Hist. Philos. Sci. A* **16**, 171–201 (1985).

Ismael, J. & Schaffer, J. Quantum holism: nonseparability as common ground. *Synthese* **197**, 4131–4160 (2020).

Itano, W.M. et al. *Quantum Zeno effect*. Phys. Rev. A **41**, 2295 (1990).

Katz, N. et al. *Coherent state evolution in a superconducting qubit from partial-collapse measurement*. Science **312**, 1498–1500 (2006).

Kennard, E. H. Zur Quantenmechanik einfacher Bewegungstypen. *Z. Phys.* **44**, 326–352 (1927).

Kocsis, S. et al. *Observing the average trajectories of single photons in a two-slit interferometer*. Science **332**, 1170–1173 (2011).

Korotkov, A.N. & Jordan, A.N. *Undoing a weak quantum measurement of a solid-state qubit*. Phys. Rev. Lett. **97**, 166805 (2006).

Kwiat, P. et al. *Interaction-free measurement*. Phys. Rev. Lett. **74**, 4763 (1995).

Lundeen, J. S. et al. *Direct measurement of the quantum wavefunction*. Nature **474**, 188–191 (2011).

Maudlin, T. *Quantum Non-Locality and Relativity: Metaphysical Intimations of Modern Physics*. 3rd edn. Wiley-Blackwell (2011).

Mermin, N. D. QBism puts the scientist back into science. *Nature* **507**, 421–423 (2014).

Muhonen, J.T. et al. *Storing quantum information for 30 seconds in a nanoelectronic device*. Nat. Nanotech. **9**, 986–991 (2014).

Penrose, R. On gravity's role in quantum state reduction. *Gen. Relativ. Gravit.* **28**, 581–600 (1996).

Piacentini, F. et al. *Measuring incompatible observables by exploiting sequential weak values*. Phys. Rev. Lett. **117**, 170402 (2016).

Planck, M. *On the Law of Distribution of Energy in the Normal Spectrum*. Ann. Phys. **309**, 553–563 (1901).

Pusey, M. F., Barrett, J. & Rudolph, T. On the reality of the quantum state. *Nat. Phys.* **8**, 475–478 (2012).

Robertson, H. P. The uncertainty principle. *Phys. Rev.* **34**, 163–164 (1929).

Rovelli, C. Relational quantum mechanics. *Int. J. Theor. Phys.* **35**, 1637–1678 (1996).

Schaffer, J. Monism: The priority of the whole. *Philos. Rev.* **119**, 31–76 (2010).

Scully, M. O., Englert, B.-G. & Walther, H. Quantum optical tests of complementarity. *Nature* **351**, 111–116 (1991).

Teller, P. Relational holism and quantum mechanics. *Br. J. Philos. Sci.* **37**, 71–81 (1986).

Tonomura, A., Endo, J., Matsuda, T., Kawasaki, T. & Ezawa, H. Demonstration of single-electron buildup of an interference pattern. *Am. J. Phys.* **57**, 117–120 (1989).

Tumulka, R. A relativistic version of the Ghirardi–Rimini–Weber model. *J. Stat. Phys.* **125**, 821–840 (2006).

Wallace, D. *The Emergent Multiverse: Quantum Theory according to the Everett Interpretation*. Oxford University Press (2012).

Zurek, W. H. Decoherence, einselection, and the quantum origins of the classical. *Rev. Mod. Phys.* **75**, 715–775 (2003).